\documentclass[%
 reprint,
 amsmath,amssymb,
 aps,
prd,
]{revtex4-2}
\usepackage[colorlinks, linkcolor=blue, anchorcolor=blue, citecolor=blue]{hyperref}
\usepackage{lineno}
\usepackage{graphicx}
\usepackage{dcolumn}
\usepackage{subfigure}
\usepackage{bm}
\newcommand{\BESIIIorcid}[1]{\href{https://orcid.org/#1}{\hspace*{0.1em}\raisebox{-0.45ex}{\includegraphics[width=1em]{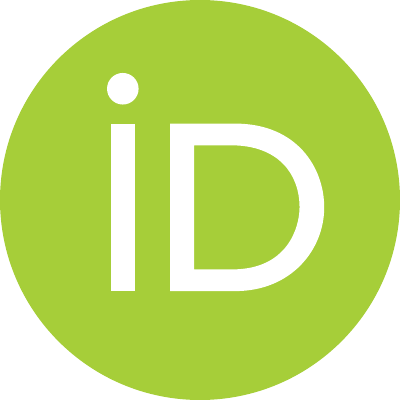}}}}

\begin{document}
\preprint{APS/123-QED}

\title{Search for a hypothetical gauge boson and dark photons in charmonium transitions}

%%%%%%%%%%%%%%%%%%%%%%%%%%%%%%%%%%%%%%%%%%%%%

\author{%% Saved at => 2025-07-02
M.~Ablikim$^{1}$\BESIIIorcid{0000-0002-3935-619X},
M.~N.~Achasov$^{4,c}$\BESIIIorcid{0000-0002-9400-8622},
P.~Adlarson$^{78}$\BESIIIorcid{0000-0001-6280-3851},
X.~C.~Ai$^{83}$\BESIIIorcid{0000-0003-3856-2415},
R.~Aliberti$^{37}$\BESIIIorcid{0000-0003-3500-4012},
A.~Amoroso$^{77A,77C}$\BESIIIorcid{0000-0002-3095-8610},
Q.~An$^{60,74,a}$,
Y.~Bai$^{59}$\BESIIIorcid{0000-0001-6593-5665},
O.~Bakina$^{38}$\BESIIIorcid{0009-0005-0719-7461},
Y.~Ban$^{48,h}$\BESIIIorcid{0000-0002-1912-0374}
% ...
\\
\vspace{0.2cm}
(BESIII Collaboration)\\
\vspace{0.2cm} {\it
$^{1}$ Institute of High Energy Physics, Beijing 100049, People's Republic of China\\
% ...
$^{4}$ Budker Institute of Nuclear Physics SB RAS (BINP), Novosibirsk 630090, Russia\\
% ...
$^{37}$ Johannes Gutenberg University of Mainz, Johann-Joachim-Becher-Weg 45, D-55099 Mainz, Germany\\
$^{38}$ Joint Institute for Nuclear Research, 141980 Dubna, Moscow region, Russia\\
% ...
$^{48}$ Peking University, Beijing 100871, People's Republic of China\\
% ...
$^{59}$ Southeast University, Nanjing 211100, People's Republic of China\\
$^{60}$ State Key Laboratory of Particle Detection and Electronics, Beijing 100049, Hefei 230026, People's Republic of China\\
% ...
$^{74}$ University of Science and Technology of China, Hefei 230026, People's Republic of China\\
% ...
$^{77}$ University of Turin and INFN, (A)University of Turin, I-10125, Turin, Italy; (B)University of Eastern Piedmont, I-15121, Alessandria, Italy; (C)INFN, I-10125, Turin, Italy\\
$^{78}$ Uppsala University, Box 516, SE-75120 Uppsala, Sweden\\
% ...
$^{83}$ Zhengzhou University, Zhengzhou 450001, People's Republic of China\\
\vspace{0.2cm}
$^{a}$ Deceased\\
% ...
$^{c}$ Also at the Novosibirsk State University, Novosibirsk, 630090, Russia\\
% ...
$^{h}$ Also at State Key Laboratory of Nuclear Physics and Technology, Peking University, Beijing 100871, People's Republic of China\\
% ...
}}
%% ends here %%

\date{\today}

\begin{abstract}
We report a direct search for a new gauge boson, $X$, with a mass of $17~\text{MeV}/c^2$, which could explain the anomalous excess of $e^+e^-$ pairs observed in the $^8\text{Be}$ nuclear transitions. The search is conducted in the charmonium decays $\chi_{cJ}\to X J/\psi~(J=0,1,2)$ via the radiative transition $\psi(3686)\to\gamma\chi_{cJ}$ using $\left(2712.4\pm 14.3 \right)\times 10^6$ $\psi(3686)$ events collected with the BESIII detector at the BEPCII collider. No significant signal is observed, and the updated upper limit on the coupling strength of charm quark and the new gauge boson, $\epsilon_c$, is set to be $|\epsilon_c|<1.2\times 10^{-2}$ at $90\%$ confidence level. We also report new constraints on the mixing strength $\epsilon$ between the Standard Model photon and dark photon $\gamma^\prime$ in the mass range from $5~\text{MeV}/c^2$ to $300~\text{MeV}/c^2$. The upper limits at $90\%$ confidence level vary within $(2.5-17.5)\times 10^{-3}$ depending on the $\gamma^\prime $ mass. 
\end{abstract}

\maketitle

\section{INTRODUCTION}

Several experiments conducted with the ATOMKI pair spectrometer, have observed anomalous excesses in the distributions of opening angles between $e^+e^-$ pairs from $^8\text{Be}$, $^4\text{He}$ and $^{12}\text{C}$ nuclear transitions~\cite{Krasznahorkay:2015iga,Krasznahorkay:2021joi,Krasznahorkay:2022pxs}. These anomalies have been interpreted as a new particle with a mass of about $17~\text{MeV}/c^2$, called the $X$ (or $X17$) boson. Many theoretical explanations for the nature of $X$ have been proposed, including protophobic vector gauge boson~\cite{Feng:2016jff,Feng:2016ysn,Feng:2020mbt}, axial vector boson~\cite{Kahn:2016vjr,Kozaczuk:2016nma}, and QCD axion~\cite{Alves:2017avw,Alves:2020xhf}, etc. Among those, the protophobic gauge boson theory proposes that $X$ couples to the Standard Model (SM) fermions with nonuniversal fermion charges $\epsilon_f$ in the units of $e$, through the Lagrangian term $\sum_f e\epsilon_f X^\mu J_\mu$, where $X^\mu$ is the gauge field of $X$, and $J_\mu=\bar\psi_f\gamma_\mu\psi_f$ is the current of ordinary SM matter. The experimental constraints on the first-generation fermion charges are also discussed. Experimentally, the NA64~\cite{NA64:2018lsq,NA64:2019auh} and NA62~\cite{NA62:2023rvm} experiments found no evidence for the existence of $X$ in the bremsstrahlung reactions and kaon decays, while the PADME collaboration recently reported an excess of events over the predicted background with local and global significance of $2.5\sigma$ and $1.8\sigma$ for an $X$ mass of $16.9~\text{MeV}/c^2$ from $e^+e^-$ annihilation processes~\cite{PADME:2025dla}. Extensive efforts are currently underway~\cite{REDTOP:2022slw,Hostert:2023tkg,Dutta:2024yjp,MEGII:2024urz,Gustavino:2024wgb,DiLuzio:2025ojt,Tran:2025fhb,Cederkall:2025bka} to study the “$17~\text{MeV}/c^2$ anomaly” and explore new physics. The nature of $X$ has triggered great interest in the physics community (see Refs.~\cite{Alves:2023ree,Gustavino:2024des} for a brief review). Therefore, it is desirable to extend the searches for the $X$ boson to $e^+e^-$ collision experiments, especially in charmonium transitions, by examining the final state $e^+e^-$ invariant mass spectrum~\cite{Liang:2016ffe,Jiang:2018uhs,BESIII:2022mxl}.

Another impetus for the search of light gauge bosons is triggered by the dark photon model. Many new physics models beyond the SM propose a dark sector to explain the dark matter, where dark and ordinary sectors interact through various portals \cite{Arkani-Hamed:2008hhe,Essig:2013lka}. One scenario is to extend the SM with an extra $U(1)$ gauge group, which leads to a neutral spin $1$ mediator called the dark photon $\gamma^\prime$~\cite{Fayet:1980ad,Holdom:1985ag}. The dark photon couples to the SM photon through the kinetic mixing term $\epsilon F^{\prime\mu\nu}F_{\mu\nu}/2$, where $F^{\prime\mu\nu}$ and $F_{\mu\nu}$ are the field strengths of the dark photon and the SM photon, respectively, and $\epsilon$ is the mixing strength. The dark photon model is the simplified case of the protophobic gauge boson theory, if the fermion coupling strengths $\epsilon_f$ are universal to all fermions; then $\epsilon_f$ are equal to the mixing strength $\epsilon$. Several mechanisms suggest that the mixing strength is expected to be of order $10^{-2}-10^{-6}$~\cite{Pospelov:2008jd,Essig:2009nc}. The light dark photon with a mass less than $1~\text{GeV}/c^2$, which leads to a negligible width $\ll 1~\text{MeV}$~\cite{Bjorken:2009mm,Batell:2009yf,Reece:2009un}, is of great interest in the experimental search for narrow resonances. In experiments, dark photons in the sub-GeV mass range have been extensively searched for in fixed-target~\cite{A1:2011yso,Merkel:2014avp,APEX:2011dww,HADES:2013nab,NA482:2015wmo,NA62:2019meo,NA64:2023wbi,WASA-at-COSY:2013zom}, beam dump~\cite{Bross:1989mp,Riordan:1987aw}, and collider~\cite{BaBar:2014zli,BESIII:2017fwv,BESIII:2018qzg,BESIII:2018aao,Anastasi:2015qla,KLOE-2:2018kqf,OPAL:2002vhf,PHENIX:2014duq} experiments. The $e^+e^-$ colliders provide an ideal environment to search for low mass dark photons in the sub-$\text{GeV}$ region as they can have a good understanding of backgrounds and therefore have the ability to set constraints on $\epsilon$~\cite{Essig:2009nc}.

In this paper, we report a direct search for a hypothetical gauge boson $X$ and dark photons at BESIII, in the charmonium decays $\chi_{cJ}\to \gamma^\prime(X) J/\psi~(J=0,1,2)$ via the radiative transition $\psi(3686)\to\gamma\chi_{cJ}$, where the $\gamma^\prime (X)$ boson decays into an $e^+e^-$ pair. This analysis is based on $\left(2712.4\pm 14.3 \right)\times 10^6$ $\psi(3686)$ events collected at the center-of-mass energy of $\sqrt{s}=3.686~\text{GeV}$~\cite{BESIII:2024lks} with the BESIII detector in 2009, 2012 and 2021. This is the first search for the $X$ particle and dark photons in $\chi_{cJ}$ transitions.

\section{BESIII DETECTOR AND MONTE CARLO
 SIMULATION}
 
The BESIII detector~\cite{BESIII:2009fln} records symmetric $e^+e^-$ collisions provided by the BEPCII storage ring~\cite{Yu:2016cof} in the center-of-mass energy range from $1.84$ to $4.95~\text{GeV}$, with a peak luminosity of $1.1 \times 10^{33}~\text{cm}^{-2}\text{s}^{-1}$ achieved at $\sqrt{s} = 3.773~\text{GeV}$. BESIII has collected large data samples in this energy region~\cite{BESIII:2020nme,Lu:2020imt,Zhang:2022bdc}. The cylindrical core of the BESIII detector covers $93\%$ of the full solid angle and consists of a helium-based multilayer drift chamber~(MDC), a time-of-flight system~(TOF), and a CsI(Tl) electromagnetic calorimeter~(EMC), which are all enclosed in a superconducting solenoidal magnet providing a $1.0~\text{T}$ magnetic field. The solenoid is supported by an octagonal flux-return yoke with resistive plate counter muon identification modules interleaved with steel. The charged-particle momentum resolution at $1~{\rm GeV}/c$ is $0.5\%$, and the ${\rm d}E/{\rm d}x$ resolution is $6\%$ for electrons from Bhabha scattering. The EMC measures photon energies with a resolution of $2.5\%$ ($5\%$) at $1~\text{GeV}$ in the barrel (end cap) region. The time resolution in the plastic scintillator TOF barrel region is $68~\text{ps}$, while that in the end cap region was $110~\text{ps}$ before 2015, when the system was upgraded using multigap resistive plate chamber technology providing a time resolution of $60~\text{ps}$, which benefits $83\%$ of the data used in this analysis~\cite{Li:2017jpg,Guo:2017sjt,Cao:2020ibk}.

Monte Carlo (MC) simulated data samples produced with a {\sc geant4}-based~\cite{GEANT4:2002zbu} software package, which includes the geometric description of the BESIII detector and the detector response, are used to determine detection efficiencies and to estimate backgrounds. The simulation models the beam energy spread and initial state radiation (ISR) in the $e^+e^-$ annihilations with the generator {\sc kkmc}~\cite{Jadach:1999vf,Jadach:2000ir}. The inclusive MC sample includes the production of the $\psi(3686)$ resonance, the ISR production of the $J/\psi$, and the continuum processes incorporated in {\sc kkmc}. All particle decays are modeled with {\sc evtgen}~\cite{Lange:2001uf,Ping:2008zz} using branching fractions (BFs) either taken from the Particle Data Group (PDG)~\cite{ParticleDataGroup:2024cfk}, when available, or otherwise estimated with {\sc lundcharm}~\cite{Chen:2000tv,Yang:2014vra}. Final state radiation from charged final state particles is incorporated using the {\sc photos} package~\cite{Barberio:1990ms}.

To simulate the signal decays, the process $\psi(3686)\to\gamma \chi_{cJ}~(J=0,1,2)$ is modeled according to its helicity decay amplitudes~\cite{BESIII:2017ung}; a phase space model is used for the subsequent decays $\chi_{cJ}\to \gamma^\prime(X)  J/\psi,~\gamma^\prime(X)\to e^+e^-$, and a vector meson to a pair of leptons model ({\sc photos vll} in {\sc evtgen}) is used for $J/\psi\to\ell^+\ell^-~(\ell=e,\mu)$. The signal samples either for $X$ or for different masses of $\gamma^\prime$ are individually simulated using $1$ million events per sample, where dark photon mass $m_{\gamma^\prime}$ varies from $5~\text{MeV}/c^2$ to $300~\text{MeV}/c^2$ by a step of $5~\text{MeV}/c^2$, and in particular, $m_X$ (the mass of the $X$ boson) is set to $17~\text{MeV}/c^2$. Exclusive background samples are also generated, with a yield equivalent to $2712.4\times 10^7$ $\psi(3686)$ events. These background samples include the electromagnetic Dalitz decays $\psi(3686)\to e^+e^-\chi_{cJ}$ and $\chi_{cJ}\to e^+ e^- J/\psi$, and other processes such as $\chi_{cJ}\to\gamma J/\psi$, $\psi(3686)\to\pi^0\pi^0 J/\psi$, and $\psi(3686)\to\pi^0(\eta) J/\psi$. A semi-blind analysis procedure is performed to avoid possible bias, where the analysis strategy is fixed and validated, using the MC samples and approximately $10\%$ of the full data sample prior to analyzing the full dataset.

In addition, the data samples collected at $\sqrt{s}=3.773~\text{GeV}$ and $3.650~\text{GeV}$~\cite{Ablikim:2013ntc,BESIII:2024lbn}, corresponding to integrated luminosities of $7926.8~\text{pb}^{-1}$ and $464.2~\text{pb}^{-1}$, are used to estimate the continuum backgrounds.
    
\section{DATA ANALYSIS}

\subsection{Event selection}
The final state particles in this analysis are $\gamma e^+e^-e^+e^-$ and $\gamma e^+e^-\mu^+\mu^-$. The signal events are required to have two positive and two negative charged tracks and at least one photon. 

Charged tracks detected in the MDC are required to be within a polar angle ($\theta$) range of $|\rm{cos\theta}|<0.93$, where $\theta$ is defined with respect to the $z$-axis, which is the symmetry axis of the MDC. For charged tracks, the distance of closest approach to the interaction point (IP) must be less than 10\,cm along the $z$-axis, $|V_{z}|$, and less than 1\,cm in the transverse plane, $|V_{xy}|$. 

Photon candidates are identified using isolated showers in the EMC. The deposited energy of each shower must be more than $25~\text{MeV}$ in the barrel region ($|\cos \theta|< 0.80$) and more than $50~\text{MeV}$ in the end cap region ($0.86 <|\cos \theta|< 0.92$), where $\theta$ is the polar angle previously defined. To exclude showers that originate from charged tracks, the angle subtended by the EMC shower and the position of the closest charged track at the EMC must be greater than $10$ degrees as measured from the IP. To suppress electronic noise and showers unrelated to the event, the difference between the EMC time and the trigger start time is required to be within $[0, 700]~\text{ns}$.

Particle identification (PID) uses the measured information in the MDC, TOF, and EMC. The combined likelihoods ($\mathcal{L}_{\text{p}}$) under the electron or positron, pion, and kaon hypotheses are obtained. Electron or positron candidates are required to satisfy $\mathcal{L}_{\text{p}}(e)>0.001$ and $\mathcal{L}_{\text{p}}(e)/\left(\mathcal{L}_{\text{p}}(e)+\mathcal{L}_{\text{p}}(\pi)+\mathcal{L}_{\text{p}}(K)\right)>0.8$. Muon candidates are required to satisfy $\mathcal{L}_{\text{p}}(\mu)>0.001$, $\mathcal{L}_{\text{p}} (\mu)> \mathcal{L}_{\text{p}} (K)$ and $\mathcal{L}_{\text{p}} (\mu)>\mathcal{L}_{\text{p}}(e)$. 

If there is more than one photon candidate, the combination with the smallest energy difference $\Delta E=|E_{\mathrm{cms}}-E_{\psi(3686)}|$ is retained, where $E_{\mathrm{cms}}$ is the center-of-mass energy and $E_{\psi(3686)}$ is the energy of the reconstructed $\psi(3686)$ particle.

For the $\gamma e^+e^-e^+e^-$ final state, the highest-momentum electron and positron tracks are considered to be leptons from $J/\psi$ decay. The other two tracks are considered as leptons from $\gamma^\prime(X)$ decay. Simulation studies of this assignment show that the misidentification rate is negligible ($<1\%$). The $J/\psi$ and $\psi(3686)$ signals are identified by requiring the invariant masses of the high momentum lepton pair and the $\gamma e^+e^-\ell^+\ell^-$ candidate to be within the intervals $|M_{\ell^+\ell^-}-M_{J/\psi}|<0.05~\text{GeV}/c^2$ and $|M_{\gamma e^+e^-\ell^+\ell^-}-M_{\psi(3686)}|<0.06~\text{GeV}/c^2$. The mass windows correspond to about $\pm 3\sigma$ of their mass resolutions, where $M_{J/\psi}$ and $M_{\psi(3686)}$ are the nominal masses of $J/\psi$ and $\psi(3686)$ particles taken from the PDG~\cite{ParticleDataGroup:2024cfk}, and $\sigma$ is the resolution obtained from the fits to the $M_{\ell^+\ell^-}$ and $M_{\gamma e^+e^-\ell^+\ell^-}$ distributions of signal MC samples with a Crystal Ball function convolved with a Gaussian function.

\begin{figure}[!htb]
\renewcommand{\figurename}{Fig}
  \centering
  \includegraphics[width=8.5cm]
  {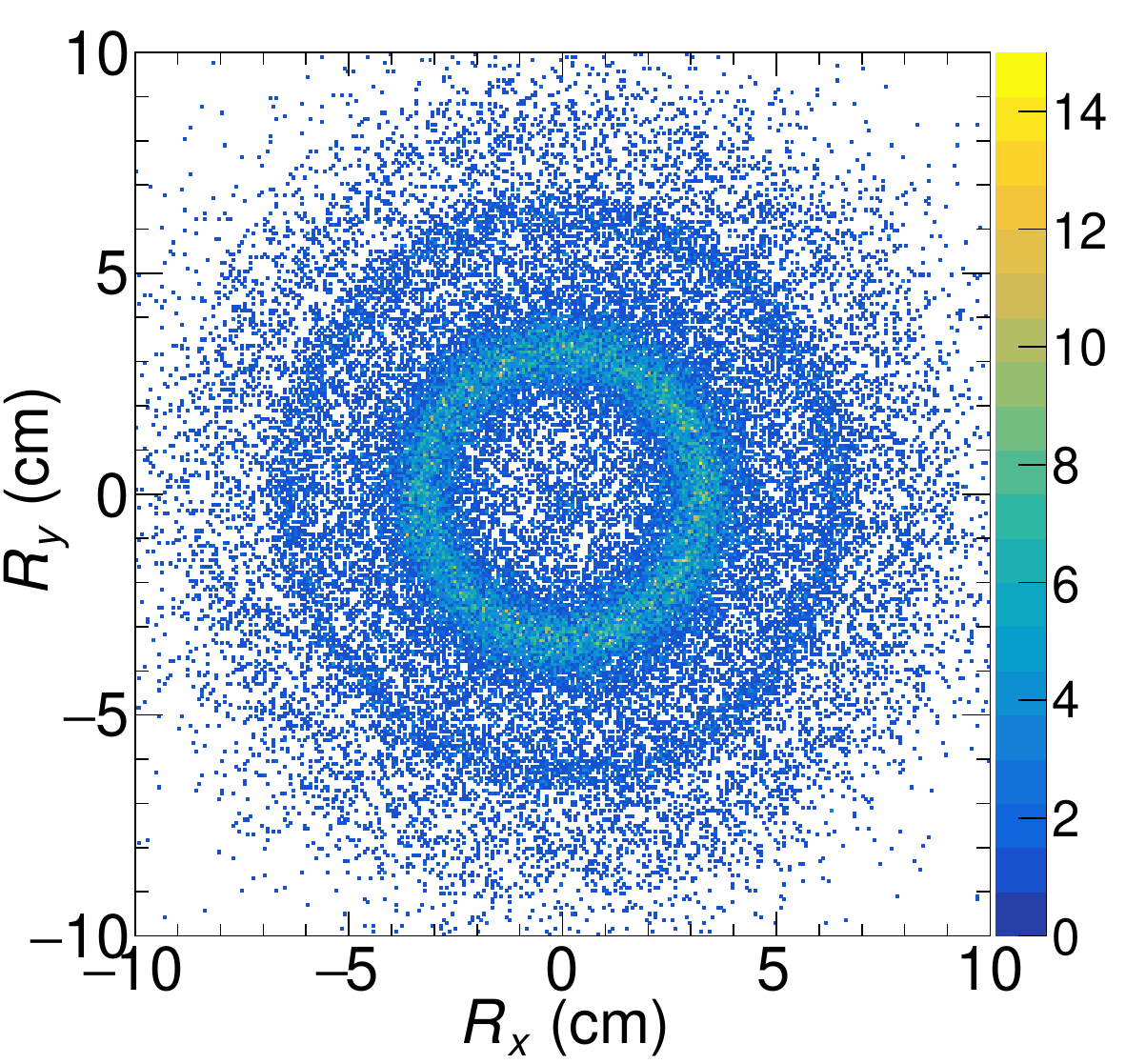}
  \caption{The distribution of the photon conversion points for simulated events from the process $\psi(3686)\to\gamma\chi_{cJ},~\chi_{cJ}\to\gamma J/\psi$. A photon converts into an electron-positron pair in the beam pipe or the inner wall of the MDC, corresponding to the inner and the outer rings.}
  \label{RXY}
\end{figure}

A study of the inclusive MC sample, using the TopoAna~\cite{Zhou:2020ksj} tool, shows that the main background arises from $\psi(3686)\to\gamma\chi_{cJ},~\chi_{cJ}\to\gamma J/\psi$ events, where one photon converts to an $e^+e^-$ pair in the detector material. To suppress these background events, a photon conversion veto algorithm~\cite{Xu:2012xq} is applied to reconstruct the conversion vertex. The length $R_{xy}$ is defined as the projected distance from the $(0,0,0)$ point to the conversion vertex of the lower momentum $e^+e^-$ pair in the $x-y$ plane. For this kind of background, $R_{xy}$ mainly accumulates at $3~\text{cm}$ and $6~\text{cm}$, corresponding to the positions of the beam pipe and the inner wall of the MDC, respectively, as displayed in Fig.~\ref{RXY}. $R_{xy}$ is required to be less than $2~\text{cm}$ to reject this background. Simulation studies show that the rejection rates of the photon conversion veto are $(77.17\pm 0.26)\%$, $(68.40 \pm 0.17)\%$, and $(67.40\pm 0.18)\%$ for three $\chi_{cJ}$ channels.

To exclude backgrounds from $\psi(3686)\to\eta(\pi^0)J/\psi$, $\eta(\pi^0)\to\gamma e^+e^-$ decays, the invariant mass $M_{\gamma e^+e^-}$ of the photon and the lower momentum $e^+e^-$ pair is required to be outside the mass windows of $[0.115,0.150]~\text{GeV}/c^2$ and $[0.505,0.570]~\text{GeV}/c^2$, respectively.

To distinguish $\chi_{c0}$, $\chi_{c1}$, and $\chi_{c2}$ candidates, the photon energy $E_\gamma$ is required to be in the ranges of $[0.23,0.28]$, $[0.15,0.18]$ and $[0.11,0.14]~\text{GeV}$, and the $e^+e^-\ell^+\ell^-$ invariant mass $M_{e^+e^-\ell^+\ell^-}$ is required to be in the ranges of $[3.39,3.47]$, $[3.48,3.55]$ and $[3.52,3.58]~\text{GeV}/c^2$, respectively. The photon energy and $e^+e^-\ell^+\ell^-$ mass requirements are optimized by maximizing the figure-of-merit $\varepsilon/\left(a/2+B\right)$ proposed by G.~Punzi~\cite{Punzi:2003bu}. Here, $a$ stands for the signal significance assumed to be $3$; $\varepsilon$ and $B$ are the detection efficiency and the number of expected background events estimated by the signal and inclusive MC samples, respectively.

\subsection{Background analysis}

After the selections, there are two types of remaining backgrounds. The first is from $\psi(3686)$ decays, which is studied using the exclusive MC samples. The dominant contribution arises from the electromagnetic Dalitz decays $\psi(3686)\to e^+e^-\chi_{cJ}$ and $\chi_{cJ}\to e^+ e^- J/\psi$, accounting for more than $50\%$ of the total background. Other contributing processes mainly include $\chi_{cJ}\to\gamma J/\psi$, $\psi(3686)\to\pi^0\pi^0 J/\psi$, and $\psi(3686)\to\pi^0(\eta) J/\psi$. The expected yields of different backgrounds are listed in Table. \ref{BACK}. The second type of background, the continuum background, is studied using data samples collected at $\sqrt{s}=3.773~\text{GeV}$ and $3.650~\text{GeV}$, and is found to be negligible. 

\renewcommand{\tablename}{Table}
        \begin{table}[htbp]
 \caption{The expected yields of different backgrounds from $\psi(3686)$ decays for three $\chi_{cJ}$ channels.}	
 \centering
 \setlength{\tabcolsep}{11pt}{
	\begin{tabular}{lcccccc}
		\hline\hline\noalign{\smallskip}	
		Source (for $\chi_{c0}$ channel) & Yield\\
\noalign{\smallskip}\hline\noalign{\smallskip}
		$\psi(3686)\to\gamma\chi_{c0},~\chi_{c0}\to e^+e^- J/\psi$ & $247\pm 16$ \\
       $\psi(3686)\to\pi^0\pi^0 J/\psi$&$119\pm 11$\\
		$\psi(3686)\to e^+e^-\chi_{c0},~\chi_{c0}\to\gamma J/\psi$& $28\pm 5$ \\
        $\psi(3686)\to\eta J/\psi,~\eta\to \gamma e^+ e^-$ &$23\pm 5$ \\   
Others&$18\pm 4$\\
\hline\hline\noalign{\smallskip}
\hline\hline\noalign{\smallskip}
		Source (for $\chi_{c1}$ channel) & Yield\\
\noalign{\smallskip}\hline\noalign{\smallskip}
		$\psi(3686)\to\gamma\chi_{c1},~\chi_{c1}\to e^+e^- J/\psi$ & $8933\pm 95$ \\
       $\psi(3686)\to\gamma\chi_{c1},~\chi_{c1}\to \gamma J/\psi$&$292\pm 17$\\
		$\psi(3686)\to\gamma\chi_{c2},~\chi_{c2}\to e^+e^- J/\psi$& $67\pm 8$ \\
        $\psi(3686)\to\pi^0\pi^0 J/\psi$ &$48\pm 7$ \\             
         $\psi(3686)\to\eta J/\psi,~\eta\to \gamma e^+ e^-$&$17\pm 4$\\
         Others &$16\pm 4$\\
\hline\hline\noalign{\smallskip}
\hline\hline\noalign{\smallskip}
		Source (for $\chi_{c2}$ channel) & Yield\\
\noalign{\smallskip}\hline\noalign{\smallskip}
		$\psi(3686)\to\gamma\chi_{c2},~\chi_{c2}\to e^+e^- J/\psi$ & $5435\pm 74$ \\
       $\psi(3686)\to\gamma\chi_{c1},~\chi_{c1}\to e^+e^- J/\psi$&$177\pm 13$\\
		$\psi(3686)\to\gamma\chi_{c2},~\chi_{c2}\to \gamma J/\psi$& $140\pm 12$ \\
        $\psi(3686)\to\to\eta J/\psi,~\eta\to \gamma e^+ e^-$ &$33\pm 6$ \\    
         $\psi(3686)\to\pi^0\pi^0 J/\psi$ &$30\pm 5$ \\    
Others&$17\pm 4$\\
\hline\hline\noalign{\smallskip}

	\end{tabular}}
\label{BACK}
\end{table}

\subsection{Fits to the data}
\begin{figure*}[htbp]
\centering  
\subfigure{
\includegraphics[width=0.32\textwidth]{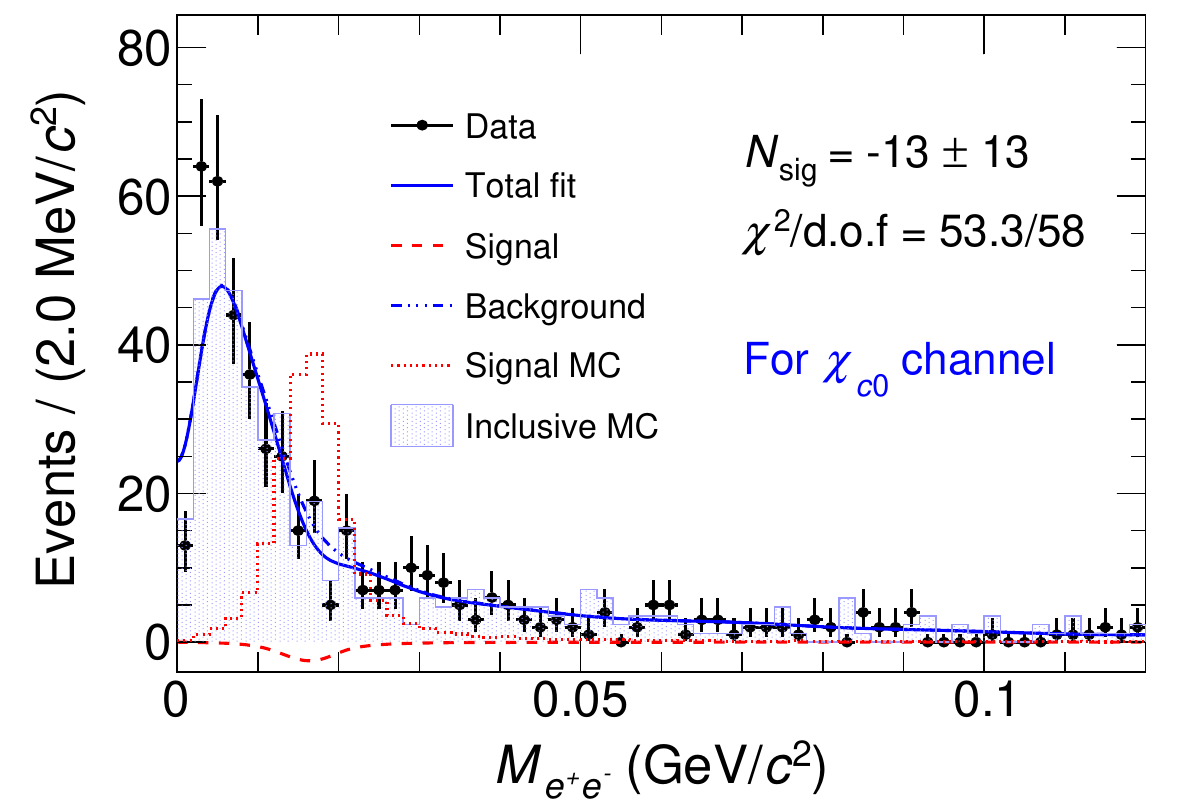}}
\subfigure{
\includegraphics[width=0.32\textwidth]{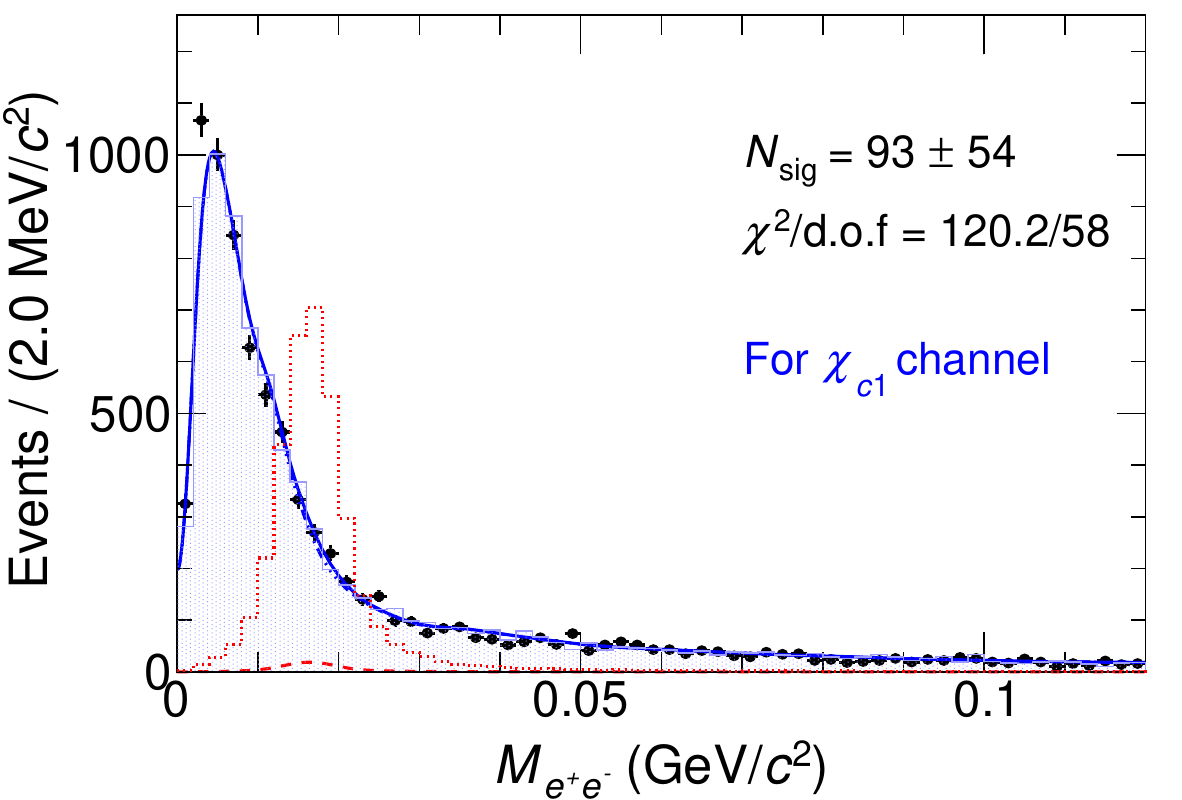}}
\subfigure{
\includegraphics[width=0.32\textwidth]{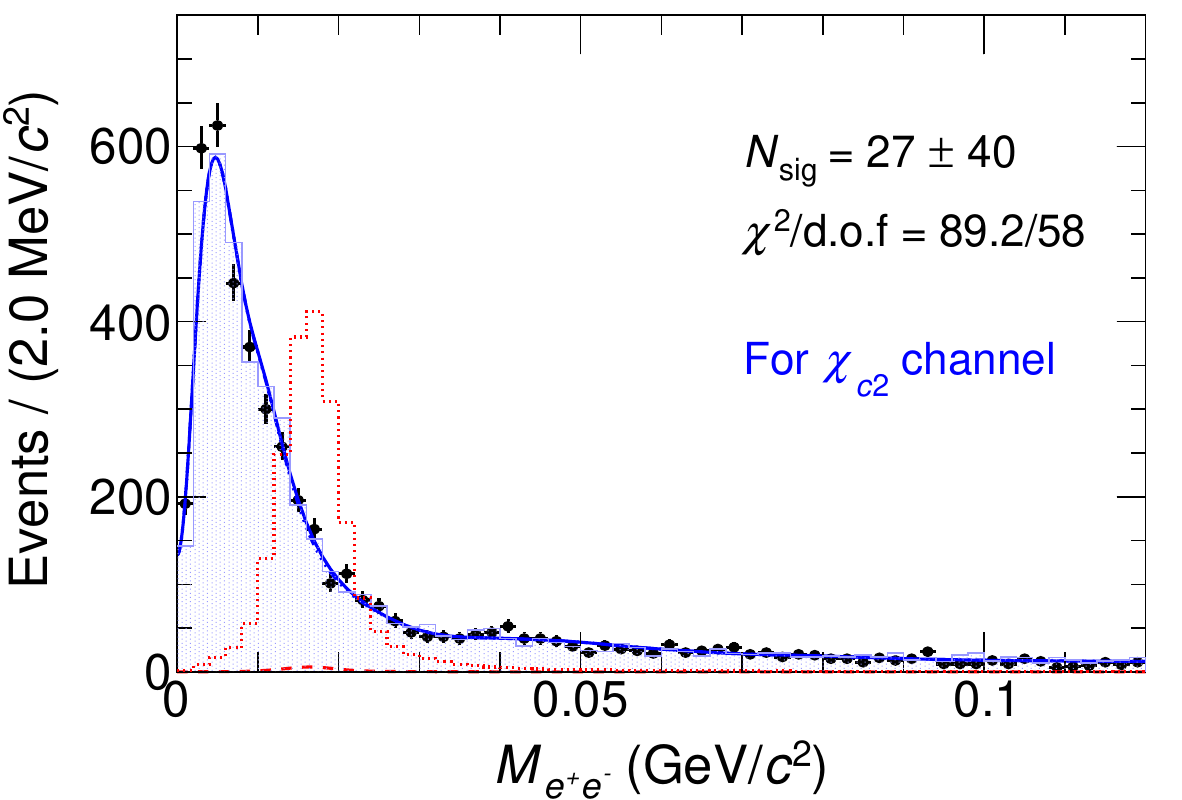}}
\caption{The fits to the $M_{e^+e^-}$ distributions of three $\chi_{cJ}$ decay channels. The dots with error bars are data samples. The blue solid curves are the total fit results, where the $\chi_{cJ}\to X J/\psi, ~X\to e^+e^-$ signals and background components are shown as the red dashed and blue dash-dotted curves, respectively. The $\chi^2/\text{d.o.f}$ is displayed on each figure as an indication of the goodness of fit, where the d.o.f is the number of degrees of freedom in each fit. The signal and inclusive MC histograms are also shown here, represented by the red dotted lines and blue shaded areas, which are scaled to half and full sizes of data, respectively.}
\label{FIT}
\end{figure*}

With the selections above, the detection efficiencies estimated by the signal MC samples of the $\chi_{c0,1,2}\to X J/\psi,~X\to e^+ e^-$ channels are $(4.39 \pm 0.02)\%$, $(7.13 \pm 0.02)\%$, and $(7.10 \pm 0.03)\%$, respectively. The signal yields, $N_{\text{sig}}$, of the three $\chi_{cJ}\to \gamma^\prime (X) J/\psi,~\gamma^\prime (X)\to e^+ e^-$ channels are determined from unbinned extended maximum likelihood fits to the distributions of the invariant mass $M_{e^+e^-}$ of the lower momentum $e^+e^-$ pair independently. The signal and background shapes are derived from the signal MC and exclusive background MC samples, respectively. The data distributions and fit results of three $\chi_{cJ}\to X J/\psi,~X\to e^+ e^-$ decay channels are shown in Fig.~\ref{FIT}. No significant signal is observed, and the corresponding statistical significance is found to be less than $2\sigma$ in each case. The statistical significance is calculated by comparing the fit likelihoods with and without the signal component.

Two checks are performed to validate this analysis. After applying the $E_\gamma$ selection, the $\chi_{cJ}\to e^+e^-J/\psi$ signals are extracted by using unbinned extended maximum likelihood fits to the $M_{e^+e^-\ell^+\ell^-}$ distributions, and the measured BFs of $\chi_{cJ}\to e^+e^-J/\psi$ are found to be consistent with the PDG values~\cite{ParticleDataGroup:2024cfk}. In addition, an input-output check is performed, which shows there is no significant bias in the fit procedures.

\renewcommand{\thesubfigure}{(\roman{subfigure})}
\renewcommand{\figurename}{Fig}

\section{SYSTEMATIC UNCERTAINTIES}

The systematic uncertainties are divided into two categories, which are multiplicative and additive uncertainties. The multiplicative uncertainties affect the detection efficiency, arising from MDC tracking, PID, photon detection, intermediate BFs, number of $\psi(3686)$, MC statistics, signal model, and selection criteria. The additive uncertainties are related to fits, coming from signal and background shapes. The multiplicative uncertainties from different sources are summarized in Table~\ref{SYS}, and described in details in the following. The total multiplicative systematic uncertainty is obtained by adding the individual components in quadrature. For additive uncertainties, alternative fits are examined, and the most conservative upper limits are taken to be the final results.

\renewcommand{\tablename}{Table}
        \begin{table}[htbp]
 \caption{Summary of the multiplicative systematic uncertainties ($\%$) for three $\chi_{cJ}$ channels.}	
 \centering
 \setlength{\tabcolsep}{11pt}{
	\begin{tabular}{ccccccc}
		\hline\hline\noalign{\smallskip}	
		Source & $\chi_{c0}$  & $\chi_{c1}$ & $\chi_{c2}$\\
\noalign{\smallskip}\hline\noalign{\smallskip}
		MDC tracking & $0.7$& $0.6$& $0.6$ \\
        PID &$4.9$&$5.4$&$5.5$\\
		Photon detection& $1.0$ &$1.0$&$1.0$\\
        Intermediate BFs &$2.4$ &$2.8$&$2.5$\\
        Number of $\psi(3686)$ events &$0.5$& $0.5$& $0.5$\\
        Signal model&$4.5$&$3.3$&$3.0$\\
        MC statistics &$0.3$&$0.3$&$0.3$\\
        $J/\psi$ mass window& $0.5$&$0.1$&$0.1$\\
        $\psi(3686)$ mass window& $0.1$&-&-\\
        Photon conversion veto&$1.0$&$1.0$&$1.0$\\
        $\pi^0(\eta)\to\gamma e^+e^-$ veto &$3.0$&$3.0$&$3.0$\\
        $\chi_{cJ}$ states identification&$2.4$&$2.6$&$3.0$ \\           
\noalign{\smallskip}\hline\noalign{\smallskip}
           Total &$8.2$&$8.1$&$8.2$\\
\hline\hline\noalign{\smallskip}
	\end{tabular}}
\label{SYS}
\end{table}

The uncertainties of the $e^\pm$ and $\mu^\pm$ tracking and PID efficiencies are studied by comparing the differences between data and MC simulations for the control samples $e^+e^-\to\gamma e^+ e^-$ and $e^+e^-\to\gamma\mu^+\mu^-$. The data-MC differences are weighted according to the distributions of the lepton momentum and $\cos\theta$ in the signal MC samples. Then the signal efficiencies applied to data are corrected by these data-MC difference factors. The uncertainties on these corrections are taken as the systematic uncertainties. The total tracking uncertainties including two low momentum $e^\pm$ tracks and two high momentum $\ell^\pm$ tracks are $0.7\%$, $0.6\%$, and $0.6\%$ for $\chi_{c0}$, $\chi_{c1}$, and $\chi_{c2}$ channels, respectively. The total PID uncertainties are $4.9\%$, $5.4\%$, and $5.5\%$ for $\chi_{c0}$, $\chi_{c1}$, and $\chi_{c2}$ channels, respectively.

The uncertainty on the photon detection is studied with a control sample of $J/\psi\to\rho^0\pi^0$ process~\cite{BESIII:2011ysp}, and is $1.0\%$ per photon. The BFs of $\psi(3686) \to \gamma \chi_{cJ}$ and $J/\psi\to \ell^+\ell^-$ are taken from the PDG~\cite{ParticleDataGroup:2024cfk}. The relative uncertainties of BFs are taken as the systematic uncertainties, which are $2.4\%$, $2.8\%$, and $2.5\%$ for $\chi_{c0}$, $\chi_{c1}$, and $\chi_{c2}$ channels, respectively. The total number of $\psi(3686)$ events has been determined with inclusive hadronic events with an uncertainty of $0.5\%$, as described in Ref.~\cite{BESIII:2024lks}. 

The uncertainties from the signal MC model are estimated by changing the signal MC generator from the phase space model to two different models, {\sc photos vll} model and a two body model ({\sc  helamp 1.0 0.0 0.0 0.0 -1.0 0.0} in {\sc evtgen}). The maximum relative changes of the signal efficiencies, which are $4.5\%$, $3.3\%$, and $3.0\%$ for $\chi_{c0}$, $\chi_{c1}$, and $\chi_{c2}$ channels, are taken as the signal model uncertainties.

The statistical uncertainty of the signal efficiency is estimated from signal MC events, which is $\Delta_\varepsilon=\sqrt{\varepsilon\left(1-\varepsilon\right)/N}$, where $N$ is the number of produced signal MC events. The relative uncertainties of the signal efficiencies $\Delta_\varepsilon/\varepsilon$ are  assigned as the systematic uncertainties due to MC statistics. 
 
The systematic uncertainties due to detection efficiencies in the $J/\psi$ and $\psi(3686)$ mass windows are studied by changing the ranges to $\pm 2.8\sigma$, $\pm 3.2\sigma$, etc. The largest relative differences on the detection efficiencies are taken as the systematic uncertainties~\cite{NA62:2022exp}, as listed in Table~\ref{SYS}. 
 
The systematic uncertainty due to the photon conversion veto is estimated to be $1.0\%$ with the control sample of $J/\psi\to\pi^+\pi^-\pi^0,~\pi^0\to\gamma e^+e^-$~\cite{BESIII:2022jde}. The systematic uncertainty due to the $\psi(3686)\to\pi^0(\eta)J/\psi,~\pi^0(\eta)\to\gamma e^+e^-$ veto is estimated to be $3.0\%$ with the control sample of $\psi(3686)\to\gamma\chi_{cJ},~\chi_{cJ}\to e^+e^- J/\psi,~J/\psi\to \ell^+\ell^-$~\cite{BESIII:2022jde}. 
 
The systematic uncertainties from different $\chi_{cJ}$ states identification ($E_\gamma$ and $M_{e^+e^-\ell^+\ell^-}$ requirements) are estimated by varying the requirements of $E_\gamma$ and $M_{e^+e^-\ell^+\ell^-}$ by $\pm 1~\text{MeV}$ or $\pm 1~\text{MeV}/c^2$, respectively. The maximum relative differences of the detection efficiencies are taken as the systematic uncertainties~\cite{NA62:2022exp}, which are $2.4\%$, $2.6\%$, and $3.0\%$ for $\chi_{c0}$, $\chi_{c1}$, and $\chi_{c2}$ channels, respectively. 
 
The systematic uncertainties associated with the signal and background shapes are evaluated by changing the interpolation order in the RooHistPdf tool~\cite{Verkerke:2003ir} within a reasonable range, and replacing background shapes obtained from different sizes of the exclusive MC samples, with a yield $9$ to $11$ times larger than the $\psi(3686)$ data sample. Among the results of these fits, the observed upper limits on the BF fluctuate slightly, and the largest one is chosen.

\section{RESULTS}

\begin{figure}[!htb]
\renewcommand{\figurename}{Fig}
      \centering
      \includegraphics[width=8.5cm]
      {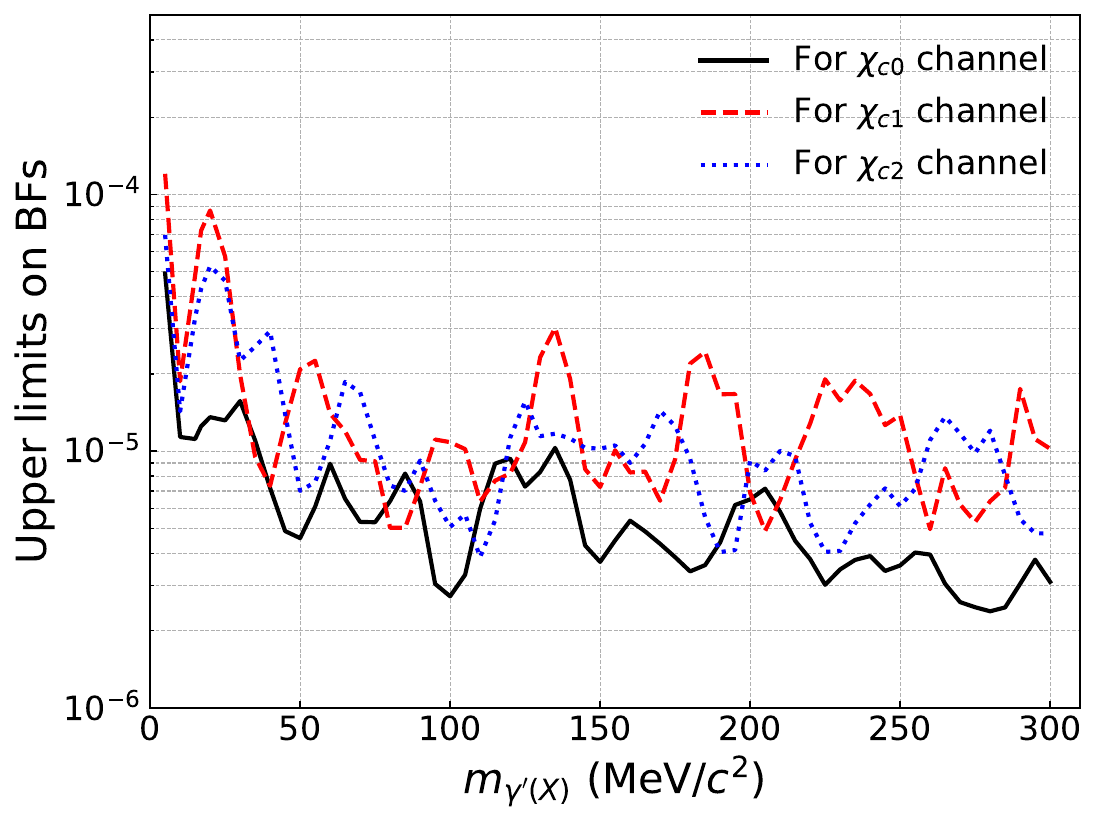}
      \caption{The ULs on BFs of $\chi_{cJ}\to \gamma^\prime(X) J/\psi,~\gamma^\prime (X)\to e^+e^-$ at $90\%$ C.L. as a function of $\gamma^\prime (X)$ mass for three $\chi_{cJ}$ channels. The black solid, red long-dashed and blue dotted curves represent $\chi_{c0}$, $\chi_{c1}$ and $\chi_{c2}$ channels, respectively.}
      \label{BF}
    \end{figure}

Since no significant signal is observed, the upper limit (UL) on the product BF is set using a Bayesian approach. The product BF of $\chi_{cJ}\to \gamma^\prime(X) J/\psi,~\gamma^\prime (X)\to e^+e^-$ is defined to be
\begin{flalign}
\mathcal{B}_{\gamma^\prime(X)}=\frac{N^{\text{UL}}}{N_{\psi(3686)}\cdot \varepsilon\cdot \mathcal{B}_{\text{inter}}},
\end{flalign}
where $N^{\text{UL}}$ is the UL of the signal yield; $N_{\psi(3686)}$ is the total number of $\psi(3686)$ events in data; $\varepsilon$ is the detection efficiency, and $\mathcal{B}_{\text{inter}}$ are the intermediate BFs, which are $(1.17\pm 0.03)\%$, $(1.16\pm 0.03)\%$, and $(1.12\pm 0.03)\%$ for $\chi_{c0}$, $\chi_{c1}$, and $\chi_{c2}$ channels, respectively. The likelihood distribution $\mathcal{L}_n$ is obtained by scanning the signal yield $n$ from $0$ to $300$ with a step size $0.01$. The relative likelihood is defined as $\mathcal{L}=\mathcal{L}_n/\mathcal{L}_{\mathrm{max}}$, where $\mathcal{L}_{\mathrm{max}}$ is the maximum likelihood value. To incorporate multiplicative systematic uncertainties, the relative likelihood distribution is smeared by a Gaussian function following the method in Ref.~\cite{Stenson:2006gwf}. The smeared likelihood is 
\begin{flalign}
\mathcal{L}^{\prime}\left(n\right)\propto\int_{0}^{1}\mathcal{L}\left(n\cdot\frac{\varepsilon ^\prime}{\varepsilon}\right)e^{-\frac{\left(\varepsilon ^\prime-\varepsilon\right)^{2}}{2\sigma^2_{\varepsilon}}}\mathrm{d}\varepsilon^\prime,
\end{flalign}
where $\sigma_{\varepsilon}=\Delta_{\text{sys}}\cdot \varepsilon$, and $\Delta_{\text{sys}}$ is the total systematic uncertainty. By integrating the smeared likelihood curve up to $90\%$ of the physical region $n>0$, the ULs on $\mathcal{B}_{X}$ for $\chi_{c0}$, $\chi_{c1}$, and $\chi_{c2}$ channels are set to be $1.3\times 10^{-5}$, $7.2\times 10^{-5}$, and $4.3\times 10^{-5}$ at $90\%$ confidence level (C.L.), respectively. The ULs on $\mathcal{B}_{\gamma^\prime(X)}$ depending on the $\gamma^\prime (X)$ mass for three $\chi_{cJ}$ channels are shown in Fig.~\ref{BF}.

\begin{figure}[!htb]
\renewcommand{\figurename}{Fig}
  \centering
  \includegraphics[width=8.5cm]
  {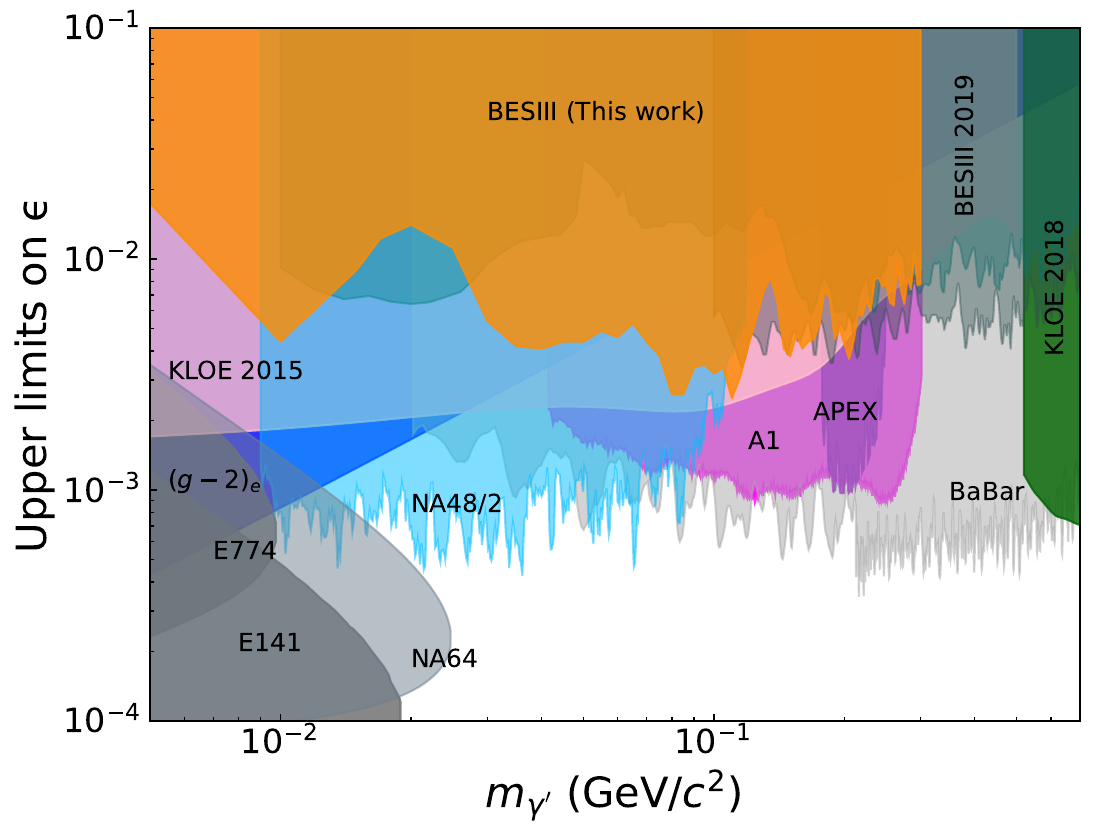}
  \caption{The $90\%$ C.L. exclusion regions of the mixing strength as a function of $\gamma^\prime$ mass from this work (orange area) and previous experiments. Bounds from the electron anomalous
magnetic moment $\left(g-2\right)_e$~\cite{Davoudiasl:2014kua} are also shown.}
  \label{DP}
\end{figure}

The mixing strength $\epsilon$ between the SM photon and the dark photon $\gamma^\prime$ is determined from the ratio of the BF $\chi_{cJ}\to \gamma^\prime  J/\psi$ and that
of the radiative process $\chi_{cJ}\to\gamma J/\psi$ as~\cite{Reece:2009un}
\begin{flalign}
    \frac{\mathcal{B}(\chi_{cJ}\to \gamma^\prime J/\psi)}{\mathcal{B}(\chi_{cJ}\to \gamma J/\psi)}=\epsilon^2 |F\left(q^2\right)|^2 \frac{\lambda^{3/2}\left(m^2_{1},m^2_{2},m^2_3\right)}{\lambda^{3/2}\left(m^2_{1},m^2_{2},0\right)},
\end{flalign}
where 
\begin{flalign}
\lambda\left(m_1^2,m_2^2,m_3^2\right)=\left(1+\frac{m_3^2}{m_1^2-m_2^2}\right)^2-\frac{4m_1^2m_3^2}{\left(m_1^2-m_2^2\right)^2},
\end{flalign}
and $m_{1,2,3}$ are the $\chi_{cJ}$, $J/\psi$ and $\gamma^\prime $ particle masses. $|F\left(q^2\right)|$ is the transition form factor from $\chi_{cJ}$ to $J/\psi$ estimated at $q^2=m^2_{\gamma^\prime }$, which is set to be $1$ according to Ref.~\cite{BESIII:2017ung}. The BF $\mathcal{B}(\chi_{cJ}\to \gamma J/\psi)$ is taken from the PDG~\cite{ParticleDataGroup:2024cfk}, and $\mathcal{B}(\gamma^\prime \to e^+e^-)$ is obtained according to Ref.~\cite{Batell:2009yf}. The ULs of the mixing strength $\epsilon$ are obtained in the combined analysis of the three independent $\chi_{cJ}$ decay channels. The likelihoods are combined via \cite{BESIII:2025ivm}
\begin{flalign}
\mathcal{L}_{\text{combine}}\left(\epsilon\right)=\mathcal{L}^\prime_{\chi_{c0}}\left(\epsilon\right)\times\mathcal{L}^\prime_{\chi_{c1}}\left(\epsilon\right)\times\mathcal{L}^\prime_{\chi_{c2}}\left(\epsilon\right),
\end{flalign}
where $\mathcal{L}^\prime_{\chi_{cJ}}\left(\epsilon\right)$ is the smeared likelihood for three $\chi_{cJ}$ channels. Then the ULs on the mixing strength $\epsilon$ at $90\%$ C.L. are evaluated by integrating the combined likelihood curve up to $90\%$ of the physical region $\epsilon>0$. Figure~\ref{DP} shows the combined exclusion limit curves on the mixing strength $\epsilon$ as a function of $\gamma^\prime $ mass from this work and the previous experiments. The ULs on the mixing strength $\epsilon$ vary within $(2.5-17.5)\times 10^{-3}$ depending on the dark photon mass. The combined upper limit on the strength of the charm quark coupling $\epsilon_c$ at $17~\text{MeV}/c^2$ is determined to be $|\epsilon_c|<1.2\times 10 ^{-2}$ using the same procedure, where the $\mathcal{B}(X\to e^+e^-)$ is assumed to be $100\%$. The sensitivity to low masses ($\sim 5~\text{MeV}/c^2$) is limited mainly due to the large backgrounds in this invariant mass region.

\section{SUMMARY}

In summary, with a sample of $\left(2712.4\pm 14.3 \right)\times 10^6$ $\psi(3686)$ events collected with the BESIII detector, the first search for a hypothetical gauge boson and dark photons in $\chi_{cJ}$ transitions is performed. No significant signal is observed. The ULs on the product BF $\mathcal{B}(\chi_{cJ}\to \gamma^\prime (X) J/\psi)\times \mathcal{B}(\gamma^\prime (X)\to e^+ e^-)$, the charm quark coupling strength $\epsilon_c$, and the mixing strength $\epsilon$ between the SM photon and dark photon are measured at $90\%$ C.L. Our results set new constraint on the $\epsilon_c$, which can help to determine the coupling between $X$ and the charm quark. Our results also improve the previous BESIII ULs on the mixing strength $\epsilon$~\cite{BESIII:2018qzg,BESIII:2018aao} for $\gamma^\prime$ masses below $0.1~\text{GeV}/c^2$. A more stringent constraint on $\epsilon$ or $\epsilon_f$ is expected in meson decay and $e^+e^-\to\gamma \gamma^\prime (X)$ process with larger data samples at the future super $\tau$-charm factory~\cite{Achasov:2023gey}.

\section*{ACKNOWLEDGEMENTS}

The BESIII Collaboration thanks the staff of BEPCII (https://cstr.cn/31109.02.BEPC) and the IHEP computing center for their strong support. This work is supported in part by National Key R\&D Program of China under Contracts Nos. 2023YFA1606000, 2023YFA1606704; National Natural Science Foundation of China (NSFC) under Contracts Nos. 11635010, 11935015, 11935016, 11935018, 12025502, 12035009, 12035013, 12061131003, 12192260, 12192261, 12192262, 12192263, 12192264, 12192265, 12221005, 12225509, 12235017, 12361141819; the Chinese Academy of Sciences (CAS) Large-Scale Scientific Facility Program; the Strategic Priority Research Program of Chinese Academy of Sciences under Contract No. XDA0480600; CAS under Contract No. YSBR-101; 100 Talents Program of CAS; The Institute of Nuclear and Particle Physics (INPAC) and Shanghai Key Laboratory for Particle Physics and Cosmology; ERC under Contract No. 758462; German Research Foundation DFG under Contract No. FOR5327; Istituto Nazionale di Fisica Nucleare, Italy; Knut and Alice Wallenberg Foundation under Contracts Nos. 2021.0174, 2021.0299; Ministry of Development of Turkey under Contract No. DPT2006K-120470; National Research Foundation of Korea under Contract No. NRF-2022R1A2C1092335; National Science and Technology fund of Mongolia; Polish National Science Centre under Contract No. 2024/53/B/ST2/00975; STFC (United Kingdom); Swedish Research Council under Contract No. 2019.04595; U. S. Department of Energy under Contract No. DE-FG02-05ER41374.

\bibliography{reference}

\end{document}